\documentclass[aps,prl,twocolumn,superscriptaddress]{revtex4-2}

\usepackage{graphicx}
\usepackage{dcolumn}
\usepackage{bm}
\usepackage{amssymb}
\usepackage{mathrsfs}
\usepackage{amsmath}
\usepackage{comment}
\usepackage{threeparttable}

\usepackage{newtxtext}

\usepackage[dvipdfmx]{hyperref}
\hypersetup{	colorlinks=true,
			linkcolor=blue,
			citecolor=blue,
			urlcolor=blue
		}	

\def\klpionn{K_L \!\to\! \pi^0 \nu \overline{\nu}}
\def\kppipnn{K^+ \!\to\! \pi^+ \nu \overline{\nu}}
\def\klpioxo{K_L \!\to\! \pi^0 X^0}
\def\klpiopiopio{K_L \!\to\! 3\pi^0}
\def\klpipipio{K_L \!\to\! \pi^+ \pi^- \pi^0}
\def\klpiopio{K_L \!\to\! 2\pi^0}
\def\klgg{K_L \!\to\! 2\gamma}

\def\kcpipio{K^{\pm} \!\to\! \pi^{\pm} \pi^0}
\def\piogg{\pi^{0} \!\to\! \gamma\gamma}
\def\to{\rightarrow}

\def\kplong{K_{L}}
\def\ppi{\pi^{0}}

\begin{document}

\title{ Search for the $\klpionn$ Decay at the J-PARC KOTO Experiment }
\newcommand{\AffChicago}{\affiliation{Enrico Fermi Institute, University of Chicago, Chicago, Illinois 60637, USA}}
\newcommand{\AffJeonbuk}{\affiliation{Division of Science Education, Jeonbuk National University, Jeonju 54896, Republic of Korea}}
\newcommand{\AffKorea}{\affiliation{Department of Physics, Korea University, Seoul 02841, Republic of Korea}}
\newcommand{\AffNTU}{\affiliation{Department of Physics, National Taiwan University, Taipei 10617, Taiwan, Republic of China}}
\newcommand{\AffKEK}{\affiliation{Institute of Particle and Nuclear Studies, High Energy Accelerator Research Organization (KEK), Tsukuba, Ibaraki 305-0801, Japan}}
\newcommand{\AffNDA}{\affiliation{Department of Applied Physics, National Defense Academy, Kanagawa 239-8686, Japan}}
\newcommand{\AffOsaka}{\affiliation{Department of Physics, Osaka University, Toyonaka, Osaka 560-0043, Japan}}
\newcommand{\AffYamagata}{\affiliation{Department of Physics, Yamagata University, Yamagata 990-8560, Japan}}
\newcommand{\AffSaga}{\affiliation{Department of Physics, Saga University, Saga 840-8502, Japan}}
\newcommand{\AffJPARC}{\affiliation{J-PARC Center, Tokai, Ibaraki 319-1195, Japan}}
\newcommand{\MarkChiba}{\altaffiliation[Present address: ]{Department of Physics and The International Center for Hadron Astrophysics, Chiba University, Chiba 263-8522, Japan.}}
\newcommand{\MarkNKNU}{\altaffiliation[Present address: ]{Department of Physics, National Kaohsiung Normal University, Kaohsiung 824, Taiwan.}}
\newcommand{\MarkNCUE}{\altaffiliation[Present address: ]{Department of Physics, National Changhua University of Education, Changhua 50007, Taiwan.}}
\newcommand{\MarkDeceased}{\altaffiliation{Deceased.}}
\author{J.~K. ~Ahn}\AffKorea
\author{M.~Farrington}\AffChicago
\author{M.~Gonzalez}\AffOsaka
\author{N.~Grethen}\AffChicago
\author{K.~Hanai}\AffOsaka
\author{N.~Hara}\AffOsaka
\author{H.~Haraguchi}\AffOsaka
\author{Y.~B. ~Hsiung}\AffNTU
\author{T. ~Inagaki}\AffKEK
\author{M.~Katayama}\AffOsaka
\author{T.~Kato}\AffOsaka
\author{Y.~Kawata}\AffOsaka
\author{E.~J. ~Kim}\AffJeonbuk
\author{H.~M. ~Kim}\AffJeonbuk
\author{A.~Kitagawa}\AffOsaka
\author{T.~K. ~Komatsubara}\AffKEK\AffJPARC
\author{K.~Kotera}\AffOsaka
\author{S.~K. ~Lee}\AffJeonbuk
\author{X.~Li}\AffChicago
\author{G.~Y. ~Lim}\AffKEK\AffJPARC
\author{C.~Lin}\MarkNCUE\AffChicago
\author{Y. ~Luo}\AffChicago
\author{T.~Mari}\AffOsaka
\author{T. ~Matsumura}\AffNDA
\author{I.~Morioka}\AffOsaka
\author{H.~Nanjo}\AffOsaka
\author{H.~Nishimiya}\AffOsaka
\author{Y.~Noichi}\AffOsaka
\author{T. ~Nomura}\AffKEK\AffJPARC
\author{K.~Ono}\AffOsaka
\author{M.~Osugi}\AffOsaka
\author{P.~Paschos}\AffChicago
\author{J.~Redeker}\AffChicago
\author{T. ~Sato}\MarkDeceased\AffKEK
\author{Y.~Sato}\AffOsaka
\author{T.~Shibata}\AffOsaka
\author{N.~Shimizu}\MarkChiba\AffOsaka
\author{T. ~Shinkawa}\AffNDA
\author{K. ~Shiomi}\AffKEK\AffJPARC
\author{R.~Shiraishi}\AffOsaka
\author{S. ~Suzuki}\AffSaga
\author{Y.~Tajima}\AffYamagata
\author{N.~Taylor}\AffOsaka
\author{Y.~C. ~Tung}\MarkNKNU\AffNTU
\author{Y.~W. ~Wah}\AffChicago
\author{H. ~Watanabe}\AffKEK\AffJPARC
\author{T. ~Wu}\AffNTU
\author{T.~Yamanaka}\AffOsaka
\author{H.~Y. ~Yoshida}\AffYamagata
\collaboration{KOTO Collaboration} \noaffiliation

\begin{abstract}
We performed a search for the $\klpionn$ decay using the data taken in 2021 at the J-PARC KOTO experiment.
With newly installed counters and new analysis method, the expected background was suppressed to  $0.252\pm0.055_{\mathrm{stat}}$$^{+0.052}_{-0.067}$$_{\mathrm{syst}}$.
With a single event sensitivity of $(9.33 \pm 0.06_{\rm stat} \pm 0.84_{\rm syst})\times 10^{-10}$, no events were observed
in the signal region. An upper limit on the branching fraction for the decay was set to be $2.2\times10^{-9}$ at the 90\% confidence level (C.L.),
which improved the previous upper limit from KOTO by a factor of 1.4. 
With the same data, a search for $\klpioxo$ was also performed, where $X^{0}$ is an invisible boson with a mass ranging from 1~MeV/$c^{2}$ to 260~MeV/$c^{2}$.
For $X^{0}$ with a mass of 135~MeV/$c^{2}$, an upper limit on the branching fraction of $\klpioxo$ was set to be $1.6\times10^{-9}$   at the 90\%~C.L.
\end{abstract}


\maketitle

\paragraph*{Introduction.} 
The KOTO experiment aims to search for the $\klpionn$ decay.
The decay directly violates CP symmetry and is highly suppressed in the Standard Model (SM) \cite{Littenberg,Kaon_Review,Brod_BEAUTY2020,DAmbrosio_2022KLpi0nunuSM}.
The SM predicts the branching fraction of this decay to be $(2.94\pm0.15)\times10^{-11}$ \cite{KLpi0nunuSM_Buras2023}, where the uncertainty
is dominated by experimental measurements of Cabibbo-Kobayashi-Maskawa matrix elements.
Because the intrinsic theoretical uncertainties are at the percent level, 
the decay is sensitive to new physics beyond the SM (e.g.,~\cite{Kpinunu_BSM_2016:1,Kpinunu_BSM_2016:2}). 
The current best experimental result is an upper limit of $3.0\times10^{-9}$ at the 90\%~C.L. \cite{KOTO2015} from the KOTO experiment \cite{KOTOproposal,KOTO}
at the Japan Proton Accelerator Research Complex (J-PARC) \cite{J-PARC} based on the data collected in 2015.
A model-independent upper limit of  $4.6\times10^{-10}$ is derived from the Grossman-Nir bound \cite{GNlimit} and the central value of the branching fraction of the $\kppipnn$ decay \cite{NA62_2018}.

In the previous analysis of KOTO of the data collected in 2016-2018, we observed three candidate events in the signal region, 
which statistically agreed with the expected number of background events $1.22\pm0.26$ \cite{KOTO2016-2018}. 
The resultant upper limit was $4.9 \times 10^{-9}$ at the 90\%~C.L.
The results of this article are based on the data taken in 2021 where several improvements were made to reduce events from the main background sources, such as $K^{\pm}$ and
beam-halo $\kplong$ backgrounds.
With the same data, new results on the search for the $\klpioxo$ decay (e.g., \cite{KLpi0X0:1,KLpi0X0:2,KLpi0X0:3,KLpi0X0:4}), where $X^{0}$ is an invisible light boson, are also reported.

\paragraph*{Experimental methods and apparatus} 

\begin{figure*}
	\includegraphics[width=1\linewidth]{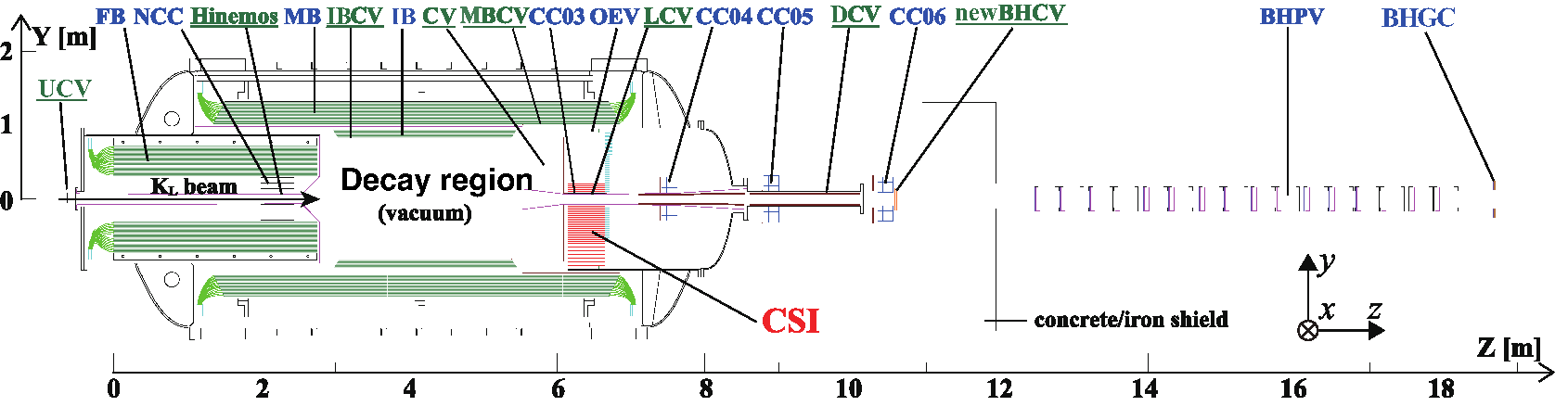}
	\caption{ 	Cross-sectional view of the KOTO detector.
			The beam enters from the left.
			Detector components with their abbreviated names written in blue regular (in green regular and underlined) are photon (charged-particle) veto counters.
		}
	\label{fig:KOTODetector}
\end{figure*}

A 30-GeV proton beam from the J-PARC main ring is extracted for a duration of 2 seconds every 5.2 seconds towards the gold target \cite{J-PARC_HEF_AuTarget2022}
located in the Hadron Experimental Facility.  
The particles produced in the target are transported into the KOTO detector at an angle of 16 degrees from the proton beam via a 20-m-long neutral beam line.
The beam line consists of two collimators and a sweeping magnet between them \cite{KOTO_BeamLine}. 
The collimators were aligned with a beam profile monitor \cite{KOTO_BPM} to minimize beam particles distributed outside the nominal beam size
due to interactions with beam line components, which are referred to as ``beam-halo" particles.
The solid angle of the beam is 7.8~{\textmu}sr, and the size is $8\times8$~cm$^{2}$ at the end of the downstream collimator.
The neutral beam is composed of $\kplong$'s, neutrons, and photons.
The flux of the $\kplong$'s, with a momentum distribution peaking at 1.4~GeV/$c$ \cite{PhD_Sato}, 
was measured to be $2.1\times10^{-7}$ $\kplong$'s per proton on target (POT) \cite{PhD_Nakagiri}.

The cross-sectional view of the KOTO detector is shown in Fig.~\ref{fig:KOTODetector}.
The $z$-axis is along the beam direction, and the origin is set at the upstream edge of FB, 21.5~m from the target.
The $\klpionn$ decay is identified by detecting the two photons from the $\ppi$ decay while observing nothing else in 
hermetic veto counters around the decay region in vacuum.
A 2-m-diameter cylindrical electromagnetic calorimeter (CSI) \cite{KOTOdet_CSI} was centered along the beam axis with a $15\times15$~cm$^2$ beam hole 
to measure the energies and positions of the two photons.
CSI was composed of 2716 undoped-CsI crystals with a length of 50~cm and a cross section of $2.5\times2.5$~cm$^{2}$ ($5\times5$~cm$^{2}$) inside
(outside) the central $1.2\times1.2$~m$^{2}$ region. In front of CSI, two layers of plastic scintillating counters (CV) \cite{KOTOdet_CV} were placed to identify charged particles
entering CSI. 
The decay region was surrounded by lead/scintillator-type sampling calorimeters, IB, MB, FB \cite{KOTOdet_IB,KOTOdet_CBARFBAR}, and OEV  \cite{KOTOdet_OEV}. 
NCC, CC03, CC04, CC05, and CC06, which were composed of undoped-CsI crystals, encircled the beam line to detect particles escaping along the beam direction.
Behind CC06,  three layers of wire chambers (newBHCV) and sixteen aerogel Cherenkov counters with lead converters (BHPV) \cite{KOTOdet_BHPV}
were placed in the beam core to detect charged particles and photons escaping through the beam hole.
The following counters were newly implemented since 2018 to further suppress several background sources.
In 2018, Multi-Pixel Photon Counters (MPPCs) were attached at the upstream surface of the undoped-CsI crystals in CSI to improve the identification of neutrons \cite{MPPC_Osugi_CHEF2019}.
A new scintillating counter (DCV) \cite{DCV_Kim_Kaon2019} was installed on the inner surface of the downstream vacuum pipe behind CSI
to detect charged particles passing through the beam hole of CSI.  
In 2021, a charged veto counter consisting of 0.5-mm-square scintillating fibers (UCV) \cite{UCV_Shiraishi_Kaon2022} was installed in the beam core
at $Z=-920~\mathrm{mm}$ to detect charged particles before the entrance of the detector.

Signals of all detector channels were recorded as waveforms by either 125-MHz \cite{KOTO_125MHzFADC} or 500-MHz digitizers  \cite{KOTO_500MHzFADC}.

\paragraph*{Data taking}
The amount of data analyzed in this article corresponds to $3.3\times10^{19}$~POT, which was collected in 2021 with a proton-beam intensity ranging from 10~kW to 64~kW.
The data acquisition system was based on two stages of the trigger logic \cite{KAON2019_Jay}.
The first level trigger (L1) required the total deposited energy in CSI to be larger than 550~MeV and no activities in NCC, MB, IB, CV,
CC03, CC04, CC05, and CC06. 
The second level trigger (L2) selected events based on the number of electromagnetic showers in CSI (nSH); 
nSH=2 was required for $\klpionn$.
Data triggered without the L2 selection was simultaneously taken with a prescale factor of 30 to collect $\klpiopiopio$, $\klpiopio$, and $\klgg$ decays
for the calculation of $\kplong$ flux.
In addition, data was collected with nSH=3 to measure the $K^{\pm}$ flux in the beam via
$\kcpipio$ decays, and with nSH=6 to estimate the beam-halo $\kplong$ flux via  $\klpiopiopio$ decays. 
 
\paragraph*{Event reconstruction and selection}
The photon energy and position in CSI were reconstructed from a cluster of adjacent crystals with
deposited energies larger than 3~MeV.  
The $\ppi$ vertex position ($Z_{\mathrm{vtx}}$) was reconstructed by assuming the $\piogg$ decay on the beam axis.
The $\ppi$ momentum perpendicular to the beam direction ($P_{t}$) was then calculated.
The $\ppi$ in $\klpionn$ decay was required to have large $P_{t}$ to account for the missing momentum taken by the two neutrinos and
to avoid the $\klpipipio$ background events where the $P_{t}$ of the $\ppi$, less than 133~$\mathrm{MeV}/c$, is limited by a kinematical boundary.
The signal region was defined in the $P_{t}$ vs $Z_{\mathrm{vtx}}$ plane as  $130 < P_{t} < 250~\mathrm{MeV}/c$ and $3200<Z_{\mathrm{vtx}}<5000~\mathrm{mm}$.
The range of $Z_{\mathrm{vtx}}$ was selected to avoid $\pi^{0}$'s generated by interactions of beam-halo neutrons with NCC and CV.
We estimated that 3.3\% of  the $\kplong$ particles from the beam exit decay in this region.
The signal region was extended to a small $P_{t}$ region compared with
the previous analysis \cite{KOTO2016-2018} because the $\klpipipio$ background was able to be further suppressed by DCV. 
To avoid bias, the event selection criteria were determined using a dataset outside the masked region enclosed by
$120 < P_{t} < 260~\mathrm{MeV}/c$ and $2900<Z_{\mathrm{vtx}}<5100~\mathrm{mm}$.

The sum of the two photon energies was required to be $>$ 650~MeV to avoid extra loss introduced by the L1 trigger.
The photon energy ($E_{\gamma}$) was required to be $100 < E_\gamma < 2000$~MeV.
To have electromagnetic showers fully contained in CSI, the photon position ($x, y$) in CSI was required to be within the fiducial region: min$(|x|,|y|) > 150$~mm and $\sqrt{x^{2}+y^{2}} < 850$~mm.
The distance between two photons was required to be $>$ 300~mm.
Consistency of their hit timing, after taking into account the time-of-flight from the reconstructed $Z_{\mathrm{vtx}}$  to CSI, was required to be within 1~ns.
The incident direction from the reconstructed $Z_{\mathrm{vtx}}$ to the photon position on CSI was required to be consistent with 
that estimated from the shape of photon clusters in CSI.
A convolutional neural network and a Fourier frequency analysis  based on cluster and pulse shapes (CSDDL and FPSD, respectively) were employed 
to eliminate clusters generated by neutrons \cite{KOTOAna_ClusterShape}.
To remove events with a pair of photons from different $\ppi$ particles in $\klpiopio$ decay, the ratio between the energies of the two photons, $E_{\gamma2}/E_{\gamma1}$ 
(where $E_{\gamma1}>E_{\gamma2}$), was required to be  $>$ 0.2.
The product of the photon energy and the photon direction to the beam axis was required to be  $>$ 2500~MeV deg.
The opening angle between the photon directions projected on the $x$-$y$ plane was required to be $<$ 150 degrees to reduce the $\klgg$ background.
$R_{\mathrm{COE}}$, defined as the distance between the beam center and the center of energy deposition in CSI, was required to be $>$ 200~mm.
The $\ppi$ kinematics was restricted on $P_{t}/P_{z}$-$Z_{\mathrm{vtx}}$ and $E$-$Z_{\mathrm{vtx}}$ planes, where $P_{z}$ and $E$ are the longitudinal momentum
and energy of the $\ppi$, respectively. 

Events were discarded if there were any hits in the veto counters coincident with the $\ppi$ decay.
Energy thresholds for the counters were set at 3.0 MeV for CC03, CC04, CC05, and CC06, 1.0~MeV for FB, NCC, MB, IB, and OEV, 
and 0.20~MeV for CV. 
Hits on newBHCV (BHPV) were required to be less than two layers (three consecutive modules) for $\klpionn$ events.
The signal acceptance and the background reduction were evaluated using Monte Carlo (MC) simulations based on GEANT4 10.5.1 \cite{GEANT4:2,GEANT4:1,GEANT4:3}.
The effect of accidental hits in detector components were taken into account in the MC simulations by overlaying waveforms recorded by a random trigger taken during physics data collection.

\begin{table}
	\caption{Summary of background estimation. The second (third) numbers represent
         the statistical uncertainties (systematic uncertainties).}
	\label{tab:BGSummary}
	\centering
	\begin{threeparttable}[h]
	\begin{tabular}{llc}
		\hline \hline
		Source & & Number of events\\
		\hline
		$K^{\pm}$ 		& 								& 0.042 $\pm$ 0.014  $^{+0.004}_{-0.028}$\\		
		$K_L$			& $\klgg$	(beam-halo)				& 0.045 $\pm$ 0.010 $\pm$ 0.006 \\
						& $\klpiopio$					        & 0.059 $\pm$ 0.022  $^{+0.050}_{-0.059}$\\								
		Neutron			& Hadron-cluster					& 0.024 $\pm$ 0.004  $\pm$ 0.006 \\
						& CV-$\eta$ 						& 0.023 $\pm$ 0.010  $\pm$ 0.005 \\
						& Upstream-$\pi^0$ 					& 0.060 $\pm$ 0.046  $\pm$ 0.007 \\
		\hline
		Total 				&								&0.252$\pm$ 0.055   $^{+0.052}_{-0.067}$\\								
		\hline \hline
	\end{tabular}
	\end{threeparttable}
\end{table}

\paragraph*{Background estimation}
Table~\ref{tab:BGSummary} summarizes the background estimation.
The total number of background events in the signal region was estimated to be $0.252\pm0.055_{\mathrm{stat}}$$^{+0.052}_{-0.067}$$_{\mathrm{syst}}$.
Background sources are categorized into three groups: $K^{\pm}$ decay backgrounds, $\kplong$ decay backgrounds, and neutron-induced backgrounds.

$K^{\pm}$'s are generated from interactions of $\kplong$ particles with the downstream collimator, after the sweeping magnet, which then may enter the
KOTO detector. This was the largest background source in the previous analysis \cite{KOTO2016-2018}.
A $\pi^{0}$ particle from $K^{\pm}$ decays may have similar kinematics to the $\pi^{0}$ generated from the $\klpionn$ decay.
The $K^{\pm}$ background was suppressed in this analysis by detecting $K^{\pm}$'s with UCV. 
The $K^{\pm}$-to-$\kplong$ flux ratio at the beam exit was evaluated to be $3.3\times10^{-5}$ by identifying the $\kcpipio$ decay with three-cluster events in CSI.
The inefficiency of UCV for detecting charged particles was measured to be $7.8^{+0.6}_{-5.2}$~\% by using the $\kcpipio$ samples,
where the main uncertainty came from the discrepancy in the low mass tail of the reconstructed $K^{\pm}$ mass.
The number of $K^{\pm}$ background events was estimated to be $0.042\pm0.014_{\mathrm{stat}}$$^{+0.004}_{-0.028}$$_{\mathrm{syst}}$,
based on the MC simulation incorporated with the UCV inefficiency obtained from data, then normalized to data using the measured $K^{\pm}$-to-$\kplong$ ratio.
The systematic uncertainties came from the $K^{\pm}$-to-$\kplong$ flux ratio and UCV inefficiency.

A beam-halo $\kplong$ that decays to 2$\gamma$ could have large reconstructed $P_{t}$
and become a background. 
The flux of beam-halo $\kplong$'s was estimated by using the MC simulations
being normalized to data by using $\klpiopiopio$ events with $R_{\mathrm{COE}}>200~\mathrm{mm}$.
To reduce this background, we implemented two selection criteria (cuts).
One is a likelihood ratio cut based on the consistency between the reconstructed photon incident angle and the resulting cluster shape in CSI.
The other is a multivariate analysis using the Fisher discrimination method to differentiate beam-halo background based on multiple kinematical variables.
They reduced the beam-halo $\klgg$ background events by a factor of 8 while keeping the signal efficiency at 94\%.  
The number of the beam-halo $\klgg$ background events was estimated to be $0.045\pm0.010_{\mathrm{stat}}\pm0.006_{\mathrm{syst}}$.
The systematic uncertainties mainly came from the cut dependence of the normalization factors for the flux of beam-halo $\kplong$'s.

The $\klpiopio$ background events were evaluated through the MC simulation by taking into account the difference of
photon detection inefficiency in the veto counters between data and the MC simulation.
We evaluated the photon detection inefficiency for veto detectors of FB, IB, MB, and BHPV by using $\klpiopiopio$
events with five out of six photons in the final state hitting on CSI.
The momentum of the remaining photon (missing photon) was then reconstructed by imposing the $\klpiopiopio$ kinematic constraints.
We defined the inefficiency as the ratio of the number of events with energy deposited in a veto detector less than 
a given threshold to the number of events having any missing photon hits in that veto detector.
The differences of the inefficiencies between data and MC were the correction factors
applied to the $\klpiopio$ background estimation, as summarized in Table~\ref{tab:2pi0_CorrectionFactor}.
We also developed a neural net cut by using kinematical variables to further suppress the $\klpiopio$ background events.
After applying the correction factors and the neural net cut, the number of $\klpiopio$ background events was evaluated to be
$0.059\pm0.022_{\mathrm{stat}}$$^{+0.050}_{-0.059}$$_{\mathrm{syst}}$. 
The systematic uncertainties came from the uncertainties on the correction factors of the photon detection inefficiency.

\begin{table}
	\caption{Correction factors on the photon detection efficiencies used in the $\klpiopio$ background estimation.}
	\label{tab:2pi0_CorrectionFactor}
	\centering
	\begin{threeparttable}[h]
	\begin{tabular}{lc}
		\hline \hline
		Detector &    
		\begin{tabular}{c}
		Difference of the inefficiency \\ between data and MC (Data/MC) 
		\end{tabular} \\
		\hline
		FB   		 					         & $1.42 \pm 0.13$\\		
		IB+MB for high energy photon\tnote{a}	&  $0.77^{+0.85}_{-0.77}$\\		
		IB+MB for low energy photon\tnote{a}       & $1.10 \pm 0.10$ \\
		BHPV							& $1.50^{+0.42}_{-0.51}$\\		
		\hline \hline
	\end{tabular}
	\begin{tablenotes}
		\item[a]  The photon detection inefficiency for IB and MB was evaluated together while
		the inefficiency was separately evaluated for high energy photons and low energy photons. 
		The reconstructed energy of the missing photon is required to be larger than 200~MeV or less than 50~MeV
		for the high energy photon or low energy photon, respectively.
		The statistical uncertainties on the five-cluster events are taken into account.		
	\end{tablenotes}

	\end{threeparttable}
\end{table}

The hadron-cluster background is caused by a beam-halo neutron which directly hits CSI and 
produces two hadronic clusters through hadronic interactions.
The timing difference ($\Delta \mathrm{T}$) between the front-end (using MPPC) and the back-end (using PMT) of each crystal
was newly implemented to measure the shower depth.
By requiring the photon-like showers to occur closer to the upstream end of the undoped-CsI crystals,
the neutron-induced cluster can be suppressed due to the difference between the short radiation length of photons and long interaction length of neutrons.
The combined rejection power of the $\Delta \mathrm{T}$, CSDDL, and FPSD cuts (neutron cuts)
was evaluated by neutron samples collected in a special run with a 3-mm-thick aluminum plate inserted in the beam at the $Z=-836$~mm
to enhance the beam-halo neutrons.
By optimizing the threshold of the neutron cuts,  the hadron-cluster backgrounds were reduced by a factor of 1.8 
compared with the previous analysis \cite{KOTO2016-2018} while keeping the signal efficiency at 75\%. 
The number of hadron-cluster background events was evaluated to be  $0.024 \pm0.004_{\mathrm{stat}}  \pm0.006_{\mathrm{syst}}$ by using neutron control samples.
The systematic uncertainties came from the uncertainties on the reduction efficiencies of the neutron cuts and the normalization between the neutron control samples and physics data.

An  $\eta$ particle produced by a beam-halo neutron interacting with CV may also become a background 
because the reconstructed $Z_{\mathrm{vtx}}$ is shifted upstream into the signal box due to the $\pi^{0}$ mass assumption during reconstruction.
The $\eta$ production in CV was estimated from MC simulation of beam-halo neutrons.
Due to low statistics of the $\eta$ production events in data, the number of events in the MC simulation was normalized to data 
by using the $\pi^{0}$ production events in CV which were identified in the region of $Z_{\mathrm{vtx}}>5100$~mm under certain selection criteria.
In order to check the reliability of this normalization method, the $\eta$-to-$\pi^{0}$ production ratio was confirmed with
data taken in a special run with a 3-mm-thick aluminum plate inserted in front of CV.  
There was a difference of 20\% in the production ratio between data and MC, which was taken into account as the systematic uncertainties.
The number of CV-$\eta$ background events was estimated to be  $0.023 \pm0.010_{\mathrm{stat}}  \pm0.005_{\mathrm{syst}}$.

A $\pi^{0}$ particle produced via the interaction of a beam-halo neutron with NCC could also become a background,
which is called the upstream $\pi^{0}$ background.
If the photon energy was mis-measured due to photo-nuclear interactions in CSI, 
the reconstructed $Z_{\mathrm{vtx}}$ could be shifted to the signal region.
The upstream $\pi^{0}$ events from the MC simulation of the beam-halo neutrons were normalized to data by using the events in the region of $Z_{\mathrm{vtx}}<$ 2900~mm 
under loose selection criteria.
The MC reproducibility of the energy mis-measurement due to photo-nuclear interactions was evaluated by using six-cluster events from $\klpiopiopio$ samples.
In order to select $\klpiopiopio$ events with photo-nuclear interactions in any of the six photons in CSI,
we required events to have $100<R_{\mathrm{COE}}<200~\mathrm{mm}$ and $|M_{6\gamma}-M_{\kplong}| >15$~$\mathrm{MeV}/c^{2}$,
where $M_{6\gamma}$ is the invariant mass of the six photons and $M_{\kplong}$ is the nominal $\kplong$ mass \cite{PDG2022}.
The number of events in MC was 2.64 times smaller than that in data.
By using this difference as the correction factor,  
the number of the upstream $\pi^{0}$ background events was estimated to be $0.060 \pm0.046_{\mathrm{stat}}  \pm0.007_{\mathrm{syst}}$.
The systematic uncertainties came from the statistical fluctuation of the correction factor and the cut dependence of the normalization factor for the $\pi^{0}$ production in NCC.

\begin{figure}
	\begin{center}
		\includegraphics[width=1\linewidth]{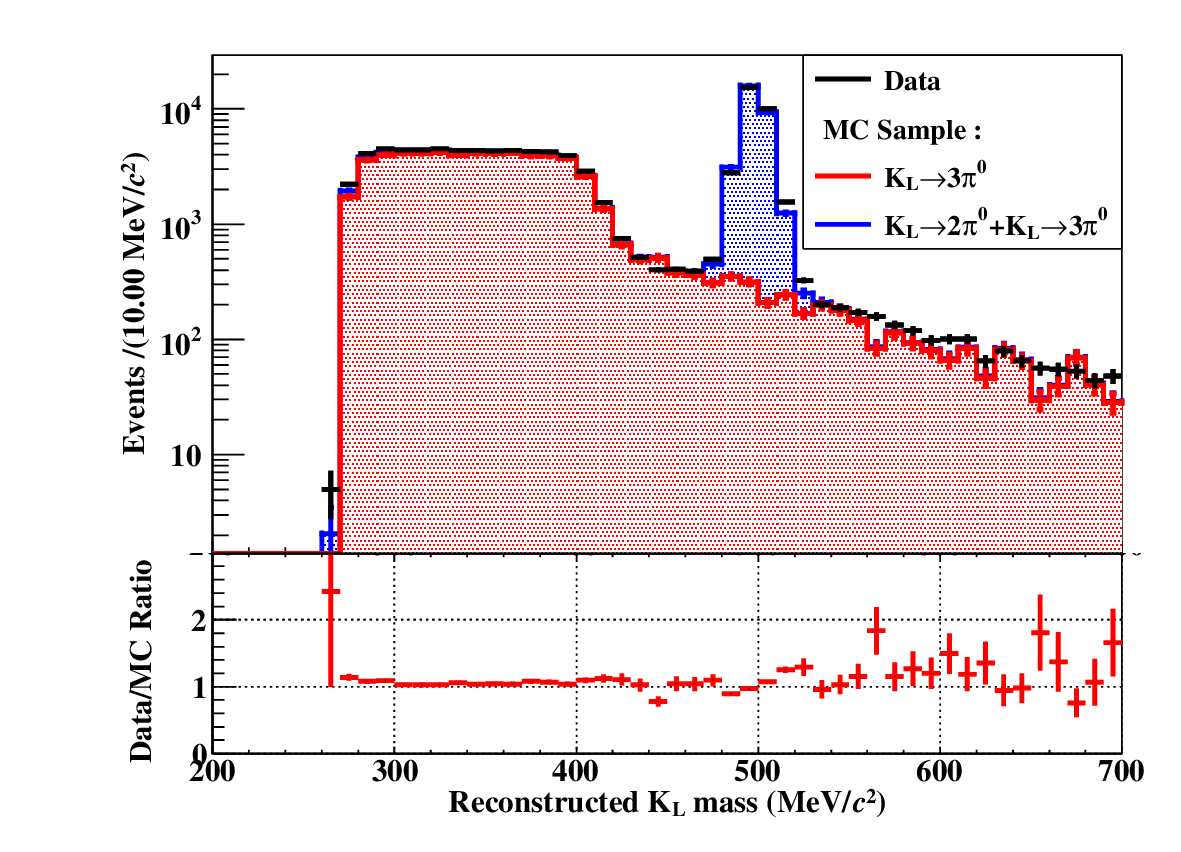}
		\caption{	
				Reconstructed $\kplong$ mass ($M_{\kplong}^{\mathrm{Rec}}$) distribution after imposing all the selection criteria except for the $M_{\kplong}^{\mathrm{Rec}}$ requirement. 
				The bottom panel shows the ratio of data to MC for each histogram bin.
				}
		\label{fig:KLMass}
	\end{center}
\end{figure}

\paragraph*{Normalization and single event sensitivity (SES)}
The SES was obtained as:

\begin{equation}
	{\rm SES} = \frac{1}{A_{\rm sig}} \frac{A_{\rm norm}\ {\rm Br}(\klpiopio)}{N_{\rm norm}},
	\label{eq:SES}
\end{equation}
where $A_{\rm norm}$ ($A_{\rm sig}$) is the acceptance of $\klpiopio$ ($\klpionn$) decays evaluated from MC simulations,
${\rm Br}(\klpiopio)$ is the nominal branching fraction for $\klpiopio$ \cite{PDG2022}, 
and $N_{\rm norm}$  is the number of reconstructed $\klpiopio$ events with a correction of the prescale factor applied to the trigger. 
Figure~\ref{fig:KLMass} shows the reconstructed $\kplong$ mass ($M_{\kplong}^{\mathrm{Rec}}$) of the $\klpiopio$ candidate events.
We required $M_{\kplong}^{\mathrm{Rec}}$ within $\pm15$~MeV/$c^{2}$ of $M_{\kplong}$. 
Based on $A_{\rm sig} =0.475$\%, $A_{\rm norm} = 0.334$\%, and $N_{\rm norm} = 6.52\times10^{5}$, the SES was estimated to be $(9.33 \pm 0.06_{\rm stat} \pm 0.84_{\rm syst})\times 10^{-10}$.

The systematic uncertainties on the SES are summarized in Table~\ref{tab:Systematics}.
The major contributions were from  the kinematical cuts for $\klpiopio$ (4.6\%), shape-related cuts (4.2\%), kinematical cuts for $\klpionn$ (3.5\%), veto cuts (3.4\%), 
and normalization modes inconsistency (3.5\%).
The uncertainties other than the last one were estimated by evaluating the discrepancy in the acceptance between data and MC for each cut and summing those differences quadratically.
A sample of $\pi^{0}$'s from the reconstructed $\klpiopio$ samples was used for the evaluation of the systematic uncertainties on shape-related cut, the kinematical cuts for $\klpionn$, and
the kinematical cuts for $\klpiopio$.
For the veto cuts,  the ratio of  the veto efficiency obtained from the reconstructed $\klpiopio$ and $\klgg$ samples was compared between data and MC.
The uncertainty on the normalization modes came from the maximum difference of the $\kplong$ flux obtained from $\klpiopiopio$, $\klpiopio$, and $\klgg$ decays.
 
\begin{table}
\caption{Relative systematic uncertainties on the single event sensitivity. The total amounts to 9.0\%.}
\label{tab:Systematics}
\centering
\begin{tabular}{lD{.}{.}{-1}}
	\hline \hline
	Source & \multicolumn{1}{c} ~Uncertainty [\%]\\
	\hline
	Kinematic cuts for $\klpiopio$ & 4.6\\
	Shape-related cuts & 4.2\\
	Normalization modes inconsistency & 3.5\\
	Kinematic cuts for $\klpionn$ & 3.5\\
	Veto cuts & 3.4\\
	Trigger effect & 1.91\\
	$K_L$ momentum spectrum & 0.98\\
	Photon selection cuts & 0.79\\        
	 $\Delta$T cut & 0.19\\
	$\klpiopio$ branching fraction & 0.69\\
	\hline
	Total &  9.0\\
	\hline \hline
\end{tabular}
\end{table}

\begin{figure}
	\begin{center}
		\includegraphics[width=1\linewidth]{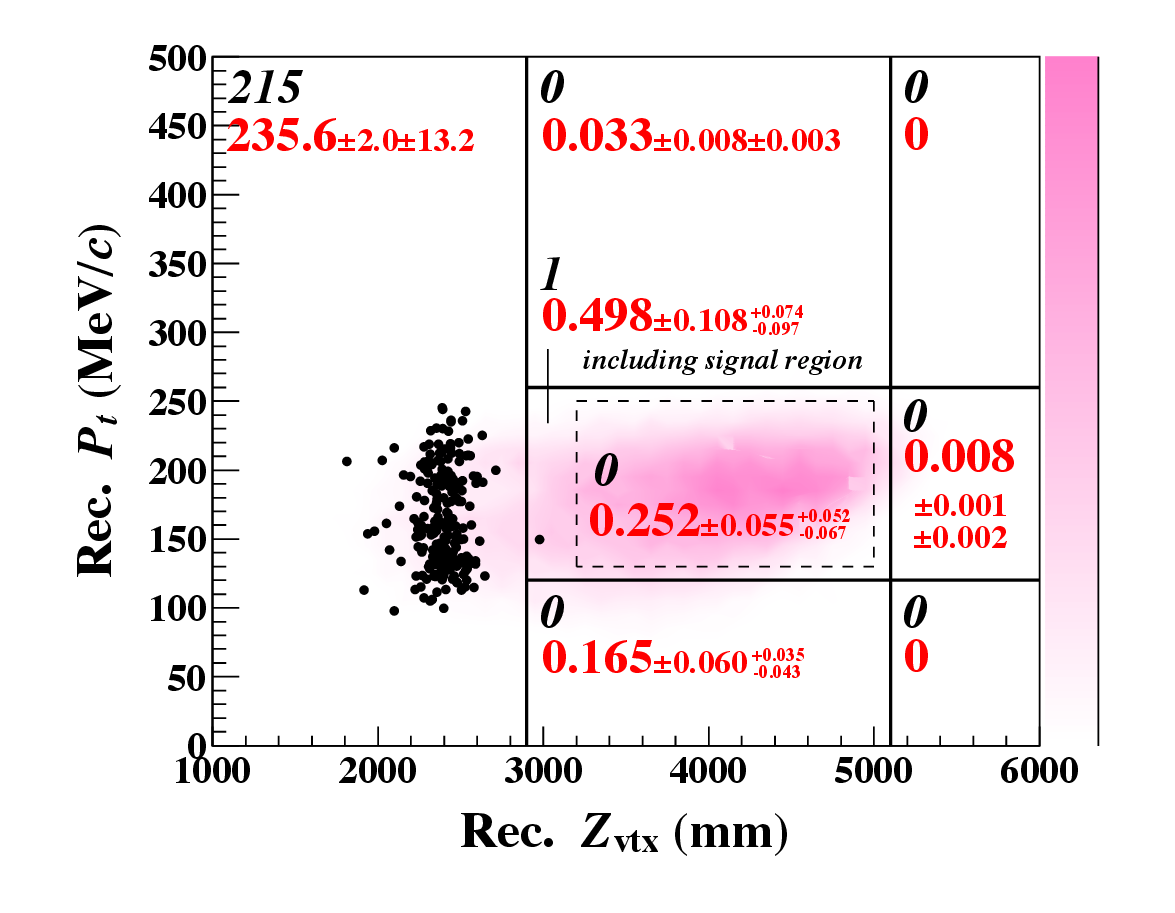}
		\caption{	
				Reconstructed $\pi^0$ transverse momentum ($P_t$) versus $\pi^0$ decay vertex position ($Z_{\rm vtx}$) plot of the events after imposing the $\klpionn$ selection criteria. 
				The region surrounded by dashed lines is the signal region. 
				The black dots represent the observed events, and the shaded contour indicates the $\klpionn$ distribution from the MC simulation. 
				The black italic (red regular) numbers indicate the number of observed (background) events for different regions.
				The second and third red regular numbers indicate the statistical uncertainties and systematic uncertainties, respectively. 				
			}
		\label{fig:FinalPtZ}
	\end{center}
\end{figure}

\begin{figure}
	\begin{center} 
           	\includegraphics[width=1\linewidth]{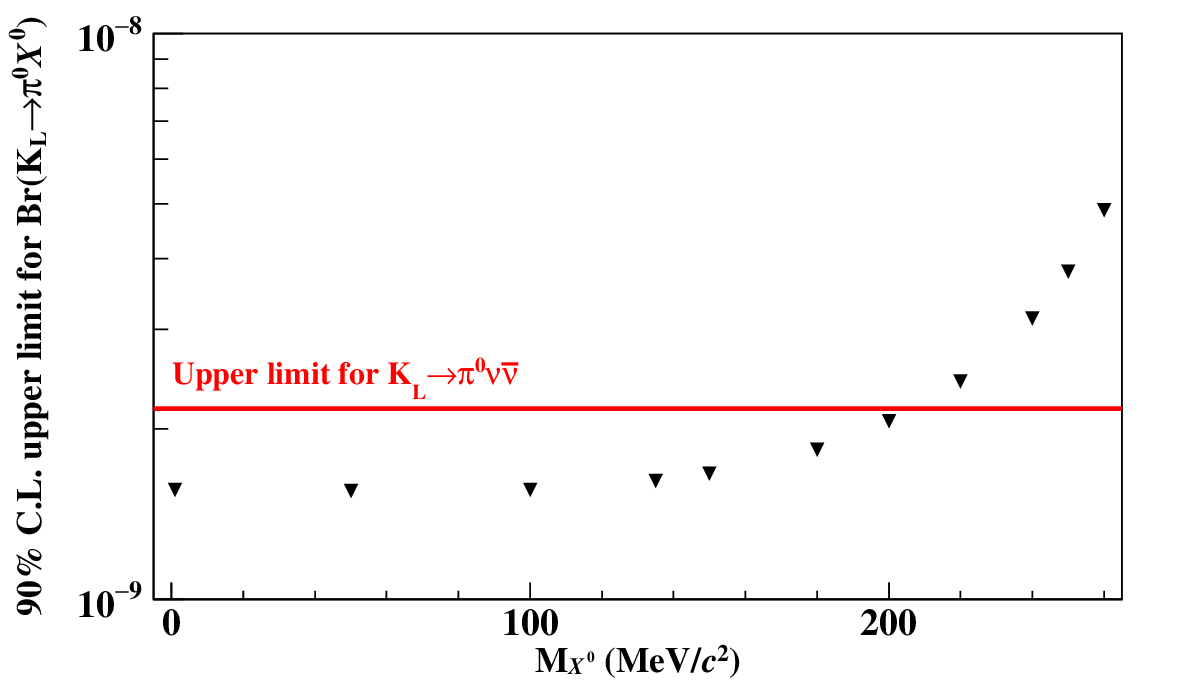}	
		\caption{	
			Upper limit at the 90\% C.L. on the branching fraction of the $\klpioxo$ decay as a function of the $X^{0}$ mass.
				}
		\label{fig:KLpi0X}
	\end{center}
\end{figure}

\paragraph*{Conclusions and prospects}
After determining all the selection criteria and estimating the number of background events,
we examined the signal region and found no events, as shown in Fig.~\ref{fig:FinalPtZ}.
By using the Poisson statistics and taking the uncertainty on SES into account \cite{UpperLimit}, we set an upper limit on the branching fraction for the $\klpionn$ decay to be 
$2.2\times10^{-9}$ at the 90\% C.L.
We also made an interpretation of the $\klpioxo$ decay with the same data. 
Figure~\ref{fig:KLpi0X} shows the 90\% C.L. upper limit on the branching fraction for $\klpioxo$ as a function of the $X^{0}$ mass ($M_{X^{0}}$); 
the limit for $M_{X^{0}} = 135~\mathrm{MeV}/c^{2}$ was set to be $1.6\times10^{-9}$ at the 90\% C.L.
These results improved the previous upper limit from KOTO by a factor of 1.4.

With this analysis, we successfully reduced the background events 
by installing new detectors and implementing new analysis methods.
By collecting more data, KOTO aims to search for the $\klpionn$ decay 
with an experimental sensitivity below $10^{-10}$.

\begin{acknowledgments}	
We would like to express our gratitude to all members of the J-PARC Accelerator and Hadron Experimental Facility groups for their support. 
We also thank the KEK Computing Research Center for KEKCC and the National Institute of Information for SINET4.
This material is based upon work supported by the Ministry of Education, Culture, Sports, Science, and Technology (MEXT) of Japan and the Japan Society for the Promotion of Science (JSPS) under the JSPS KAKENHI Grant Numbers JP16H06343, JP16H02184, 17K05480, 20K14488, 21H04483, and 21H04995, and through the Japan-U.S. Cooperative Research Program in High Energy Physics; 
the U.S. Department of Energy, Office of Science, Office of High Energy Physics, 
under Grand Number DE-SC0009798, and University of Chicago Computational Institute and the Open Science Grid Consortium;
the Ministry of Education (MOE) and the National Science and Technology Council (NSTC) in Taiwan under Grant Numbers 108-2112-M-002-001, 109-2112-M-002-021, and 110-2112-M-002-039;
and the National Research Foundation of Korea (2018R1A5A1025563, 2019R1A2C1084552, and 2022R1A5A1030700). 
Some of the authors were supported by Grants-in-Aid for JSPS Fellows.
\end{acknowledgments}	

\bibliography{bibliography}

\end{document}